\documentclass[12pt]{article}
\newcommand {\evis} {$E_{\mathrm{vis}}$}

\newcommand {\omutau} {$\nu_\mu \rightarrow \nu_\tau$}

\newcommand {\omue} {$\nu_\mu \rightarrow \nu_e$}
\newcommand {\omuebar} {$\bar{\nu}_\mu \rightarrow \bar{\nu}_e$}

\def\nue_pizero{$\nu_e n \rightarrow e^- \pi^0 p$}
\def\anue_pizero{$\bar{\nu}_e p \rightarrow e^+ \pi^0 n$}
\def\ypi_gg{$\pi^0 \rightarrow \gamma\gamma$}

\begin {document}
\title
{Multi-ring signatures of the oscillation \omue\ in a water Cherenkov 
detector}
\author{
A. Asratyan\thanks{Corresponding author. Tel.: + 7-095-237-0079;
fax: + 7-095-127-0837; {\it E-mail address:} asratyan@vitep1.itep.ru.},
G.V. Davidenko, 
A.G. Dolgolenko,
V.S. Kaftanov,\\
M.A. Kubantsev\thanks{Now at Department of Physics and Astronomy,
Northwestern University, Evanston, IL 60208, USA.},
and V. Verebryusov\\
\normalsize {\it Institute of Theoretical and Experimental Physics,}\\
\normalsize {\it B. Cheremushkinskaya St. 25, Moscow 117259, Russia}
}                                          % authors
\date {\today}
\maketitle

\begin{abstract}
     Multi-ring signatures of $\nu_e$ appearance via the oscillation 
\omue\ are formulated for a water Cherenkov detector. These signatures 
are appropriate for long-baseline neutrino experiments operating at 
relatively high neutrino energies $E_\nu > 2$ GeV that emphasize the 
matter effect. The NC background is less for selected multi-ring events 
than for 1$e$-like events, and may be directly estimated from the data. 
Sensitivity to the sign of $\Delta m^2_{31}$ and to 
$\sin^2 2\theta_{13}$ is estimated for a conceptual scheme in which the 
proposed UNO detector is exposed to a neutrino beam from Fermilab's Main 
Injector. Also discussed is the physics potential of using a water 
Cherenkov detector in the NuMI program.
\end{abstract}

     Detecting the oscillation \omue\ in long-baseline accelerator
experiments will provide clues to a number of neutrino-mixing
parameters: the mixing angle $\theta_{13}$, the sign of the 
"atmospheric" mass-squared difference $\Delta m^2_{31}$, and the 
$CP$-violating phase $\delta_{CP}$ \cite{review}.                    %{review}
The sign of $\Delta m^2_{31}$ is correlated with that of
the asymmetry between the \omue\ and \omuebar\ probabilities
induced by the MSW matter effect \cite{matter1, matter2};   %{matter1, matter2}
the magnitude of this asymmetry is roughly proportional to 
$E_\nu$ for neutrino energies well below the MSW resonance at 
some 10--15 GeV. At the same time, a similar asymmetry arising 
from $CP$ violation only depends on the ratio $L/E_\nu$.
Therefore, the two effects can be disentangled by comparing the
data for different baselines and energies. 
In the experiment JHF2K \cite{jhf2k}                                   %{jhf2k}
due to start operation by the end of the decade, Super-Kamiokande 
will be irradiated by a $\nu_\mu$ beam with 
$\langle E_\nu \rangle \simeq 0.7-0.8$ GeV over a baseline of
$L = 295$ km. Likewise, in this paper we discuss using a water 
Cherenkov detector that offers good $e/\mu$ and $e/h$ separation 
and spectrometry for electrons \cite{resolution},                 %{resolution}
but assume substantially bigger energy and baseline so as to 
emphasize the matter effect compared to JHF2K. 

\begin{figure}
\vspace{18 cm}
\includegraphics{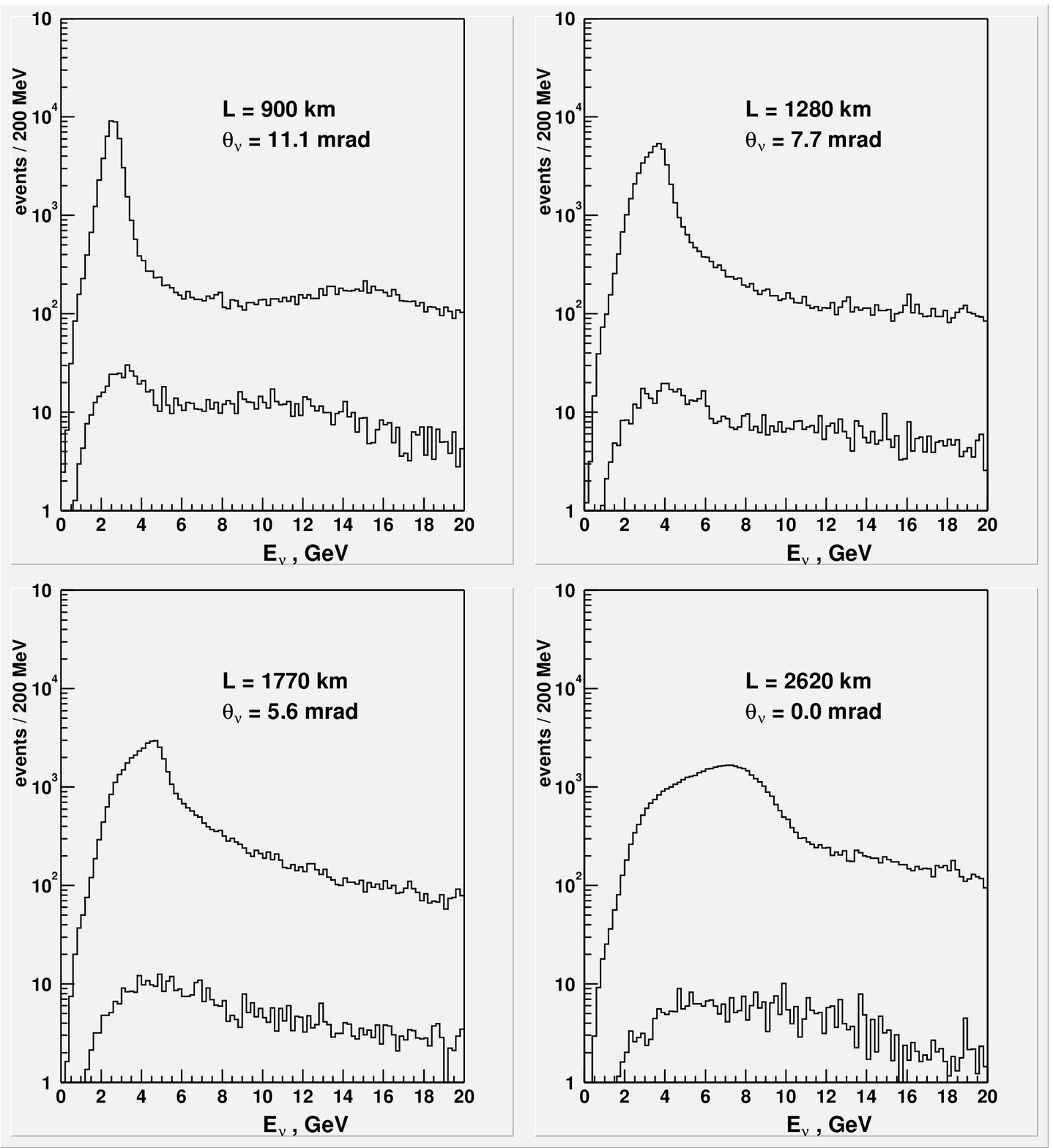}
\caption
{The oscillation-free $E_\nu$ spectra of $\nu_\mu$- and 
$\nu_e$-induced CC events (upper and lower histograms, respectively)
for four locations in the medium-energy beam of Fermilab's 
Main Injector:
$L = 900$ km and $\theta_\nu = 11.1$ mrad (top left),
$L = 1280$ km and $\theta_\nu = 7.7$ mrad (top right),
$L = 1770$ km and $\theta_\nu = 5.6$ mrad (bottom left), and
$L = 2620$ km and $\theta_\nu = 0$ mrad (bottom right).
For incident neutrinos and an exposure of 500 kton-years.}
\label{spectra}
\end{figure}

     To be specific, we assume that the medium-energy 
beam\footnote{This is taken as the PH2me beam with peak
energy of $\nu_\mu$ flux near 6 GeV at zero angle, as 
designed for the NuMI--MINOS program \cite{numi}.}                   %{numi}
of Fermilab's Main Injector is aimed at either of the three candidate 
sites for UNO---the proposed big water Cherenkov detector with fiducial 
mass of $\sim 500$ kilotons \cite{uno}. These sites include           %{uno}
the Homestake mine in Lead, South Dakota ($L = 1280$ km); the WIPP 
facility in Carlsbad, New Mexico ($L = 1770$ km); and the San Jacinto
mountain in California ($L = 2620$ km). For comparison, we also 
consider a shorter baseline of $L = 900$ km, that is near the maximum 
value of $L$ allowed by the original beamline of the NuMI--MINOS program 
\cite{numi} for far sites within 10 km of the beam axis \cite{loi}.   %{loi}
Except for the longest baseline of $L = 2620$ km, we adopt the 
concept of an "off-axis" neutrino beam \cite{offaxis}              %{offaxis}
that allows to enhance the $\nu_\mu$ flux at oscillation maximum, 
as well as to suppress the backgrounds to \omue\ arising from NC 
collisions and from the $\nu_e$ component of the original beam. The 
off-axis angle $\theta_\nu$ is selected so as to place the maximum
of the $E_\nu$ distribution near the first maximum of the 
oscillation for $|\Delta m^2_{31}| \simeq 0.003$ eV$^2$:
$\theta_\nu = 11.1$, 7.7, 5.6, and 0 mrad \cite{fluxes}            %{fluxes}
for $L = 900$, 1280, 1770, and 2620 km, respectively. 
Figure \ref{spectra} shows the oscillation-free $E_\nu$ spectra 
of all $\nu_\mu$- and $\nu_e$-induced CC events for either site, 
assuming $4\times 10^{20}$ protons on target per year \cite{numi} 
and an exposure of 500 kton--years. 

     Our foremost task is to formulate the selections of 
$\nu_e$-induced CC events appropriate for $E_\nu > 2$ GeV, so for 
simplicity the "solar" mass-squared difference $\Delta m^2_{21}$ is 
set to zero thus excluding from the simulation any effects of 
intrinsic $CP$ violation. The matter effect is estimated in
the approximation of uniform matter density
along the neutrino path \cite{matter2}, assuming                     %{matter2}
$\rho = 3$ g/cm$^3$ for all baselines considered. Relevant 
neutrino-mixing parameters are assigned the values consistent with
the atmospheric and reactor data \cite{atmo, chooz}:             %{atmo, chooz}
$\Delta m^2_{31} = \pm 0.003$ eV$^2$, $\sin^2 2\theta_{23} = 1$, and 
$\sin^2 2\theta_{13} = 0.1$ (the latter value is at the upper limit 
imposed in \cite{chooz}). The simulation relies on the               %{chooz}
neutrino-event generator NEUGEN based on the Soudan-2
Monte Carlo \cite{neugen}, that takes full account of exclusive      %{neugen}
channels like quasielastics and excitation of baryon resonances.

\begin{figure}
\vspace{18 cm}
\includegraphics{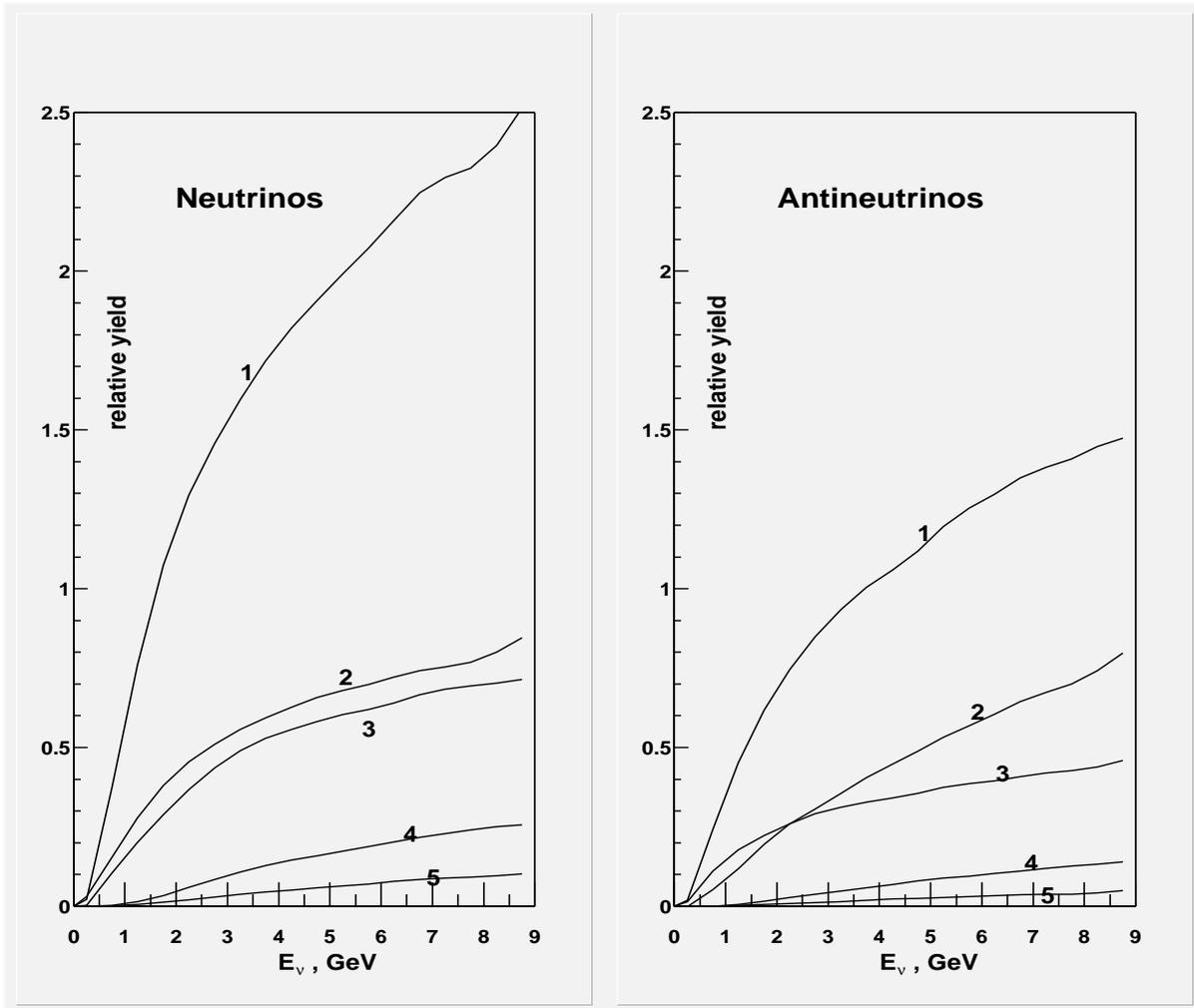}
\caption
{Cross sections per mean nucleon in water of relevant CC and NC 
reactions, divided by the $\nu_e N \rightarrow e^- X$ cross section, 
as functions of neutrino energy:
$\nu_e N \rightarrow e^- \pi^+ X$ (curve 1),
$\nu_e N \rightarrow e^- \pi^0 X$ (curve 2),
$\nu N \rightarrow \nu \pi^0 X$  (curve 3),
$\nu N \rightarrow \nu \pi^0 \pi^\pm X$ (curve 4), and
$\nu N \rightarrow \nu \pi^0 \pi^0 X$ (curve 5).
The left- and right-hand panels are for incident neutrinos and
antineutrinos, respectively.}
\label{xsections}
\end{figure}

\begin{figure}
\vspace{18 cm}
\includegraphics{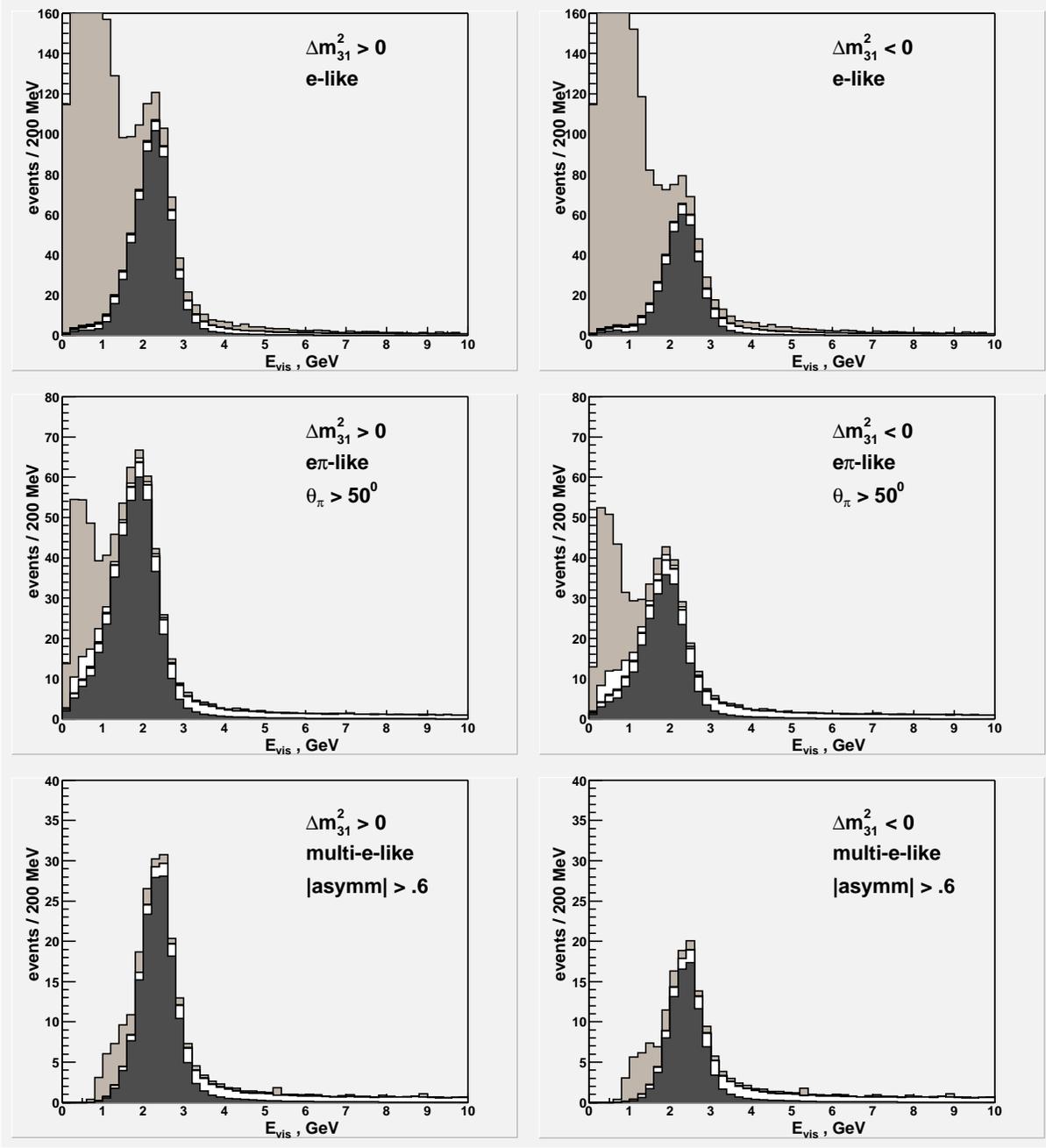}
\caption
{For $L = 900$ km, $\theta_\nu = 11.1$ mrad,
and incident neutrinos, \evis\ distributions of 
1$e$-like events (top panels), $e\pi$-like events (middle panels),
and multi-$e$-like events (bottom panels). The left- and right-hand 
panels are for $\Delta m^2_{31} > 0$ and  $\Delta m^2_{31} < 0$, 
respectively. From bottom, the depicted components are the \omue\ 
signal (shaded area), intrinsic $\nu_e$CC background (white area), 
$\nu_\tau$CC background (black area), $\nu_\mu$CC background (white
area), and the NC background (light-shaded area).}
\label{nu900all}
\end{figure}

\begin{figure}
\vspace{18 cm}
\includegraphics{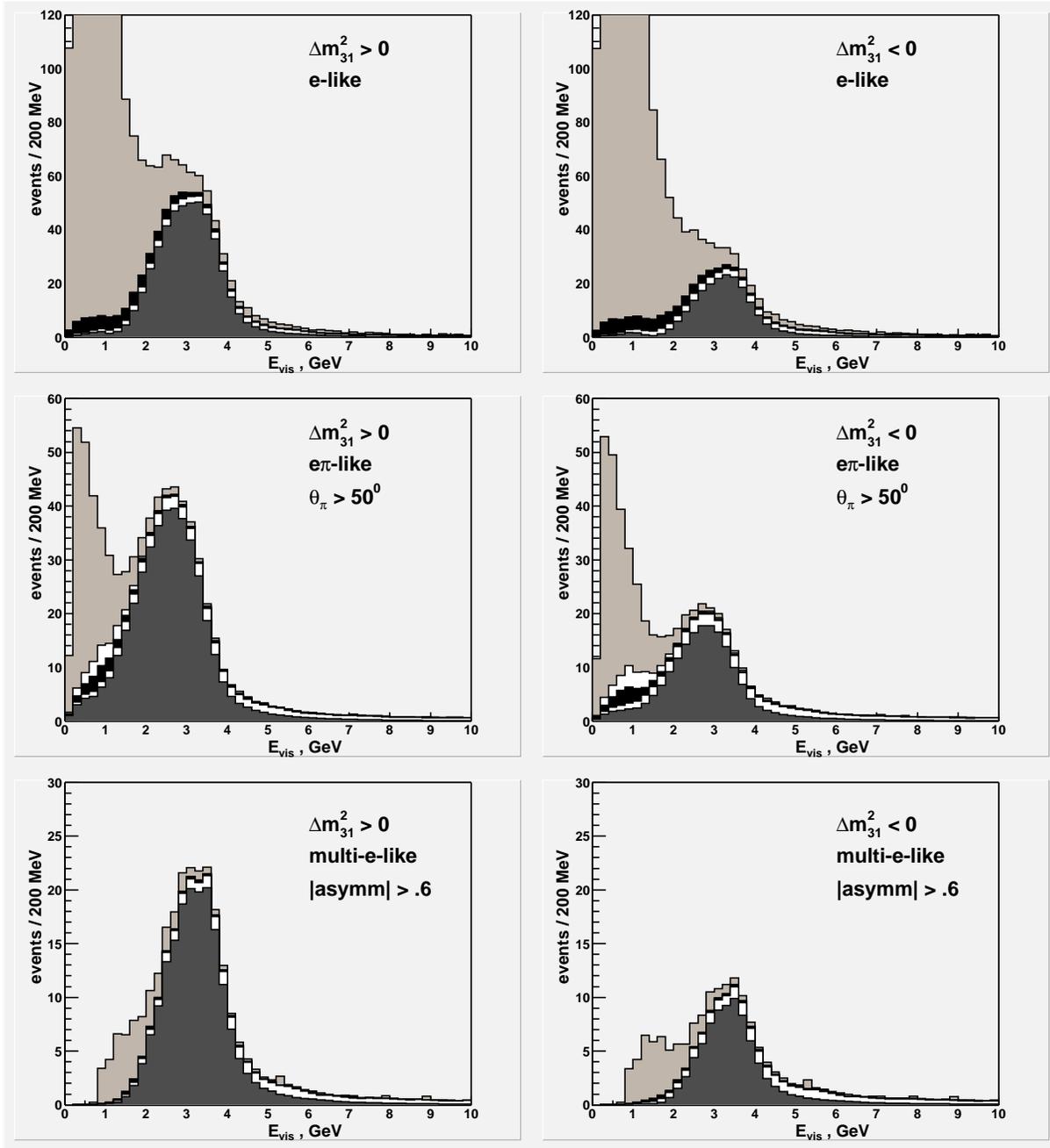}
\caption
{Same as Fig. \ref{nu900all}, but for $L = 1280$ km and 
$\theta_\nu = 7.7$ mrad.}
\label{nu1280all}
\end{figure}

\begin{figure}
\vspace{18 cm}
\includegraphics{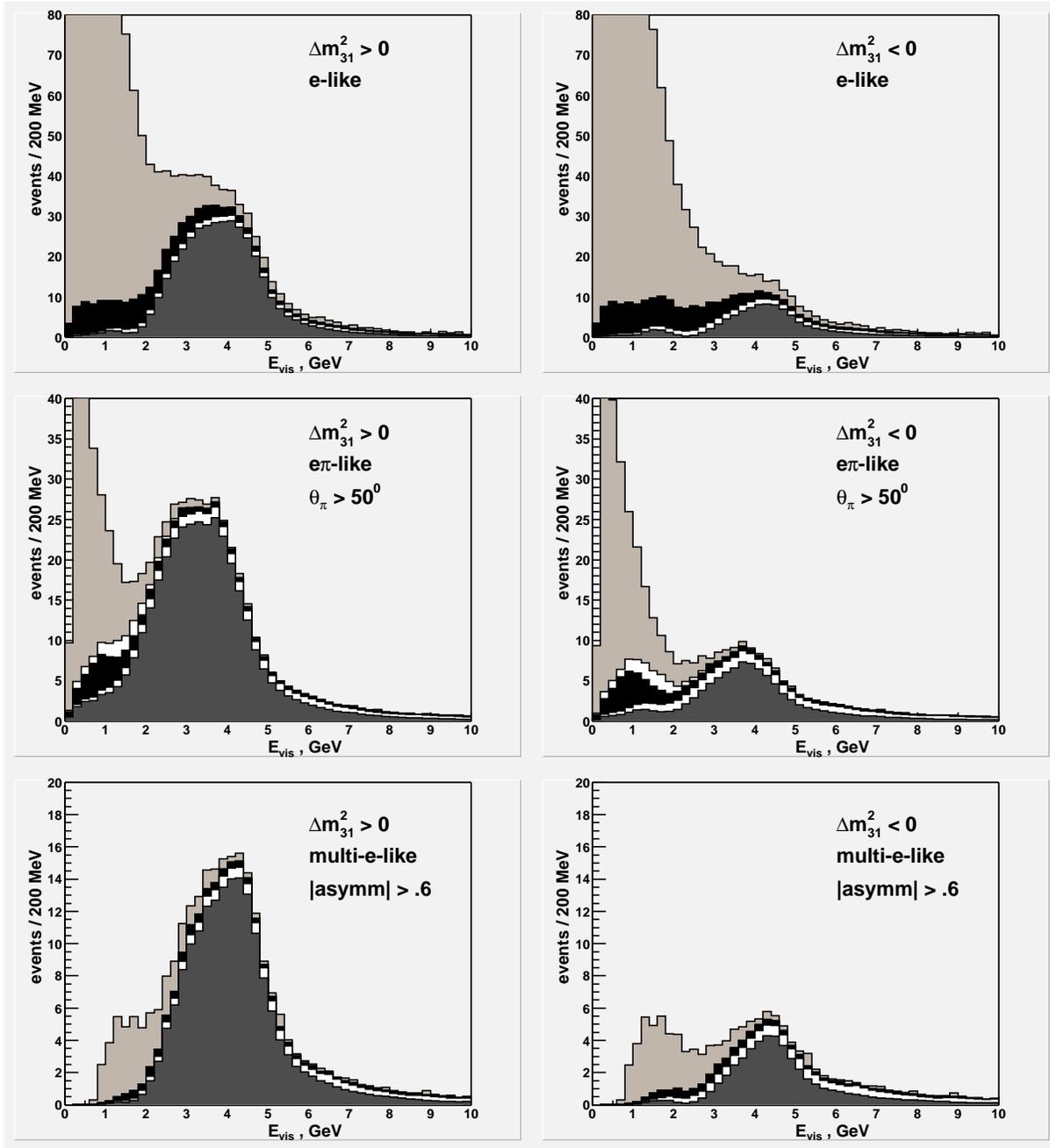}
\caption
{Same as Fig. \ref{nu900all}, but for $L = 1770$ km and
$\theta_\nu = 5.6$ mrad.}
\label{nu1770all}
\end{figure}

\begin{figure}
\vspace{18 cm}
\includegraphics{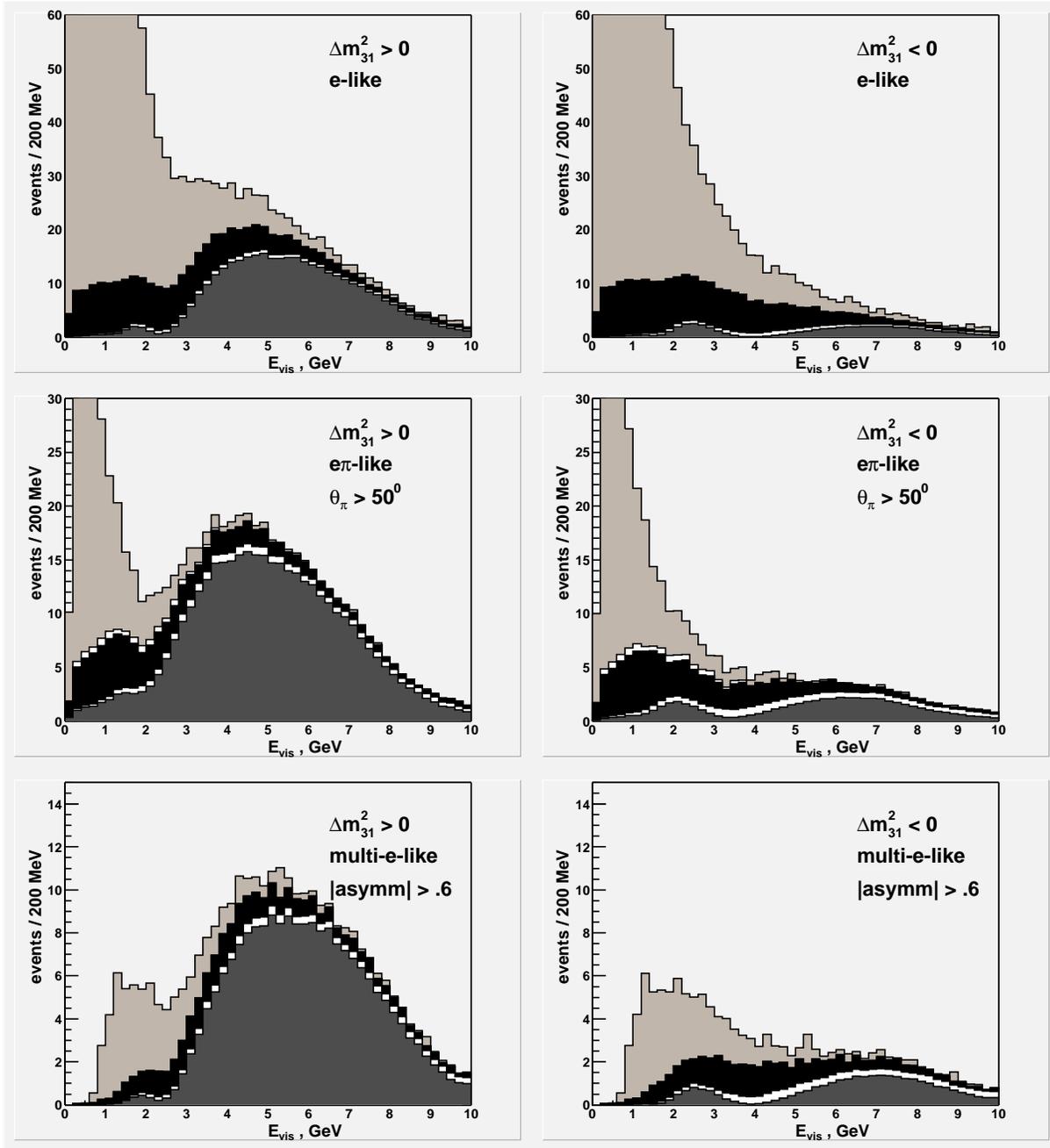}
\caption
{Same as Fig. \ref{nu900all}, but for $L = 2620$ km and
$\theta_\nu = 0$.}
\label{nu2620all}
\end{figure}

     At neutrino energies below 1 GeV, $\nu_e$ appearance can be
efficiently detected by selecting 1-ring $e$-like events of the
reaction  $\nu_e N \rightarrow e^-X$  that is dominated by 
quasielastics. (Here and in what follows, $X$ denotes a system of
hadrons other than the $\pi^0$, in which the momenta of all charged
particles are below the Cherenkov threshold in water.) The background 
largely comes from the flavor-blind NC reaction 
$\nu N \rightarrow \nu \pi^0 X$ whose cross section relative to
$\nu_e N \rightarrow e^-X$  increases with $E_\nu$, see
Fig. \ref{xsections}. At low neutrino energies $\sim 1$ GeV, this
NC reaction is suppressed by limited phase space and, moreover, the 
bulk of $\pi^0$ mesons are identified by resolving the rings of 
two photons from \ypi_gg \footnote{The efficiency 
of $\pi^0$ reconstruction, as measured in the near 
detector of the K2K experiment \cite{k2k}, steeply                     %(k2k}
decreases with increasing $\pi^0$ momentum and vanishes 
at $p(\pi^0) \simeq 900$ MeV \cite{pizero}.}.                         %{pizero}
As a result, in JHF2K the $\nu N \rightarrow \nu \pi^0 N$ background
is not expected to exceed the "intrinsic" $\nu_e$CC background due 
to the original $\nu_e$ component of the beam \cite{jhf2k}.           %{jhf2k}
The $\nu_\tau$CC background, arising from the dominant oscillation 
\omutau\ followed by  $\nu_\tau N \rightarrow \tau^-X$  and  
$\tau^- \rightarrow e^- \nu \bar{\nu}$, is negligibly small due 
to the threshold effect in $\tau$ production. At higher neutrino 
energies discussed in this paper, the NC reaction  
$\nu N \rightarrow \nu \pi^0 X$ emerges as the dominant source of
1-ring $e$-like events, and the $\nu_\tau$CC background tends to 
exceed the intrinsic $\nu_e$CC background. This is illustrated by 
the upper panels in Figs. \ref{nu900all}--\ref{nu2620all} where 
\evis\ distributions of 1$e$-like events are shown for each baseline 
and either sign of $\Delta m^2_{31}$. (Depending on the reaction, 
\evis\ stands for either the $e^-$ or $\pi^0$ energy.) These 
distributions also illustrate the dependence of matter effect on 
neutrino energy.

     As expected, the 1$e$-like signature of $\nu_e$ appearance proves
to be less rewarding for $E_\nu > 2$ GeV than for $E_\nu < 1$ GeV. In
this paper, we propose to detect the oscillation \omue\ by selecting
2- and 3-ring signatures of the reactions
$\nu_e N \rightarrow e^- \pi^+ X$  and
$\nu_e N \rightarrow e^- \pi^0 X$
that involve emission of a charged or neutral 
pion\footnote{Here and below, corresponding
antineutrino reactions are implicitly
included.}.
The motivation is that, at neutrino energies of a few GeV, the cross 
section for formation of $\Delta(1232)$ states alone is comparable to 
quasielastics, whereas the background final states $\nu \pi^0 \pi N$ 
are suppressed with respect to  $\nu \pi^0 N$. In other
words, one may expect that demanding an extra pion in the final state 
will effectively reduce the NC background rather than the CC signal.
This is illustrated by Fig. \ref{xsections} showing the cross sections
of relevant CC and NC reactions relative to $\nu_e N \rightarrow e^- X$ 
as functions of neutrino energy. The ratios between the cross sections
of corresponding CC and NC reactions,
$\sigma (\nu_e N \rightarrow e^- \pi^+ X) /
\sigma (\nu N \rightarrow \nu \pi^0 \pi^\pm X)$ and
$\sigma (\nu_e N \rightarrow e^- \pi^0 X) /
\sigma (\nu N \rightarrow \nu \pi^0 \pi^0 X)$,
are seen to be substantially larger than the ratio
$\sigma(\nu_e N\rightarrow e^-X)/\sigma(\nu N \rightarrow\nu\pi^0 X)$.

     The reaction  $\nu_e N \rightarrow e^- \pi^+ X$  will produce
two rings in the detector, of which one is $e$-like and the other is
non-showering. Apart from the reaction
$\nu N \rightarrow \nu \pi^0 \pi^\pm X$ with two pions in the final 
state, a potentially dangerous source of NC background is the more
frequent process $\nu p \rightarrow \nu \pi^0 p$ in which 
the momentum of the final proton is above the Cherenkov threshold. 
Note however that $\nu N$ kinematics restrict the emission angles of 
such protons to the region $\cos \theta > 0.45$, whereas pions with
momenta above the Cherenkov threshold may even travel in the backward
hemisphere. A lower cut on emission angle of the 
non-showering particle, $\theta > 50^0$, rejects the bulk of visible
protons from $\nu p \rightarrow \nu \pi^0 p$ and keeps nearly a half
of visible pions emitted in the reaction 
$\nu_e N \rightarrow e^- \pi^+ X$.
The latter cut will also prevent the $\pi^+$ ring from collapsing 
into the $e^-$ ring. Therefore, we select 2-ring events featuring a 
$e$-like ring and an additional ring due to a non-showering particle 
with a large emission angle of $\theta > 50^0$. This is referred to as 
the $e \pi$ signature in Table \ref{statistics} below. The distributions 
of thus selected events in visible energy \evis, defined as the energy 
of the $e$-like ring, are shown in the middle panels of 
Figs. \ref{nu900all}--\ref{nu2620all} 
for $L = 900$, 1280, 1770, and 2620 km assuming incident neutrinos. The
NC background is seen to be less for the selected $e\pi$-like events 
than for 1$e$-like events (compare with the top panels of the same
Figures). That the $\nu_\tau$CC background is also less for $e\pi$-like 
events than for quasielastics is due to a stronger 
threshold suppression of $\nu_\tau N \rightarrow \tau^- \pi^+ N$  
compared to $\nu_\tau n \rightarrow \tau^- p$. On the other hand, the 
$\nu_\mu$CC background is negligibly small for 1$e$-like events, but
contributes to selected $e\pi$-like events through the reaction
$\nu_\mu N \rightarrow \mu^- \pi^0 X$ in which the muon is emitted 
at a broad angle.

     Next, we consider the reaction $\nu_e N \rightarrow e^- \pi^0 X$ 
that features a neutral pion in the 
final state\footnote{The first observation 
in a water Cherenkov detector of the 
corresponding $\nu_\mu$-induced reaction,
$\nu_\mu N \rightarrow \mu^- \pi^0 X$,
has been reported in \cite{first}.}.                                %{first}
Depending on whether or not the $\pi^0$ has been reconstructed, 
the observable signature is either three $e$-like rings of which two 
fit to \ypi_gg, or two $e$-like rings that would not fit to a $\pi^0$.
The NC background arises from the reaction
$\nu N \rightarrow \nu \pi^0 \pi^0 N$ in which 
at least one of the two $\pi^0$ mesons has not been reconstructed.
Note that in the latter reaction the two $\pi^0$ mesons are emitted 
with comparable energies, whereas in  $\nu_e N \rightarrow e^- \pi^0 X$  
the $e^-$ tends to be the leading particle. This suggests a selection
based on the absolute value of asymmetry  $A = (E_1-E_2)/(E_1+E_2)$,
where $E_1$ and $E_2$ are the energies of the two showers for the
two-ring signature, and of the reconstructed $\pi^0$ and the "odd"
shower---for the three-ring signature. The selection $|A| > 0.6$
has been adopted in this paper. The distributions of events featuring
2 or 3 $e$-like rings in visible energy \evis, defined as the total 
energy of all such rings, are shown in the bottom panels of
Figs. \ref{nu900all}--\ref{nu2620all} for incident neutrinos. Again,
the NC and $\nu_\tau$CC backgrounds are seen to be less than for
1$e$-like events.

     The total number of \omue\ events is listed in 
Table \ref{statistics} for either sign of $\Delta m^2_{31}$, signature, 
baseline, and beam setting, assuming an exposure of 500 kton-years. Note 
that for $\Delta m^2_{31} > 0$ ($< 0$) and incident (anti)neutrinos, the 
$e\pi$ and multi-$e$ signals of \omue\ are virtually independent of
the baseline despite the depletion of neutrino flux with distance. This 
is a combined effect of matter-induced enhancement, $E_\nu$-dependence 
of one-pion production (see Fig. \ref{xsections}),
and $\theta_\nu$-dependence of neutrino flux (see Fig. \ref{spectra}).
We estimate the significance of \omue\ signals for the 1$e$,
$e\pi$, and multi-$e$ signatures in the approximation of zero 
systematic uncertainty on the background. The signal interval of
\evis\ is selected so as to maximize the quantity $F = S / \sqrt{B}$,
where $S$ and $B$ are the numbers of signal and background events
falling within the interval. Thus obtained values of $F$ are also listed
in Table \ref{statistics}.

\begin{table}[h]
\begin{tabular}{|c|c|c|c|c|c|}
\hline
$L$, $\theta_\nu$  &Signature &Total signal, &Total signal, &$F$ value, 
&$F$ value,\\
&   &$\nu$ beam  &$\bar{\nu}$ beam&$\nu$ beam &$\bar{\nu}$ beam\\
\hline
900 km, & 1$e$       & 575 (330) &195(335)  & 41.6 (25.1) & 23.8 (39.9) \\
11.1 mrad& $e\pi$     & 400 (234) &87 (146)  & 43.4 (25.4) & 17.4 (29.3) \\
        & multi-$e$  & 152 (90)  &35 (57)   & 30.4 (18.2) & 11.3 (19.1) \\
\hline
1280 km,& 1$e$       & 494 (207) &140 (328) & 34.8 (15.9) & 16.7 (36.8) \\
7.7 mrad& $e\pi$     & 414 (179) &74 (167)  & 47.7 (21.1) & 14.9 (33.1) \\
        & multi-$e$  & 184 (81)  &40 (86)   & 32.1 (14.7) & 10.4 (22.8) \\
\hline
1770 km,& 1$e$       & 383 (100) &72 (273)  & 27.9 (8.0)  & 8.6 (30.3) \\
5.6 mrad& $e\pi$     & 358 (96)  &42 (153)  & 43.8 (12.1) & 8.4 (30.2) \\
        & multi-$e$  & 174 (48)  &27 (94)   & 29.5 (8.4)  & 6.7 (23.1) \\
\hline
2620 km,& 1$e$       & 345 (54)  &40 (267)  & 23.3 (3.7)  & 4.2 (25.5) \\
0.0 mrad& $e\pi$     & 372 (57)  &26 (172)  & 39.8 (6.7)  & 4.8 (27.8) \\
        & multi-$e$  & 199 (29)  &20 (130)  & 27.8 (4.6)  & 4.4 (24.1) \\
\hline
\end{tabular}
\caption
{Total number of \omue\ events and the "figure of merit" 
$F = S /\sqrt{B}$ for either sign of $\Delta m^2_{31}$, signature, 
baseline, and beam setting, assuming $4\times10^{20}$ protons per 
year and an exposure of 500 kton--years. The first (second) value 
corresponds to the positive (negative) sign of the mass-squared 
difference $\Delta m^2_{31}$.}
\label{statistics}
\end{table}

     The NC and $\nu_\tau$CC backgrounds are dominant sources of 
systematics, and therefore we do not include the 1$e$-like events in our 
estimates of statistical significance of the \omue\ signal. Assuming 
$\Delta m^2_{31} > 0$ and incident neutrinos, for the combined sample of 
$e\pi$-like and multi-$e$-like events we obtain 
$F$ = 52.8, 57.4, 52.7, and 48.5 for $L$ = 900, 1280, 1770, and 2620 km,
respectively. (Note that the significance of the \omue\ signal, like
its magnitude, is
fairly independent of the baseline.) Thus at 90\% C.L., a one-year 
exposure of UNO in the $\nu$ beam will allow to probe the value of 
$\sin^2 2\theta_{13}$ down to 0.0022--0.0026 (depending on the baseline).
Assuming $\Delta m^2_{31} < 0$, after a one-year exposure of UNO 
in the $\bar{\nu}$ beam the experiment will be sensitive to 
$\sin^2 2\theta_{13}$ values down to 0.0032--0.0034. As soon as the 
number of delivered protons is increased from $4\times 10^{20}$
to $1.6\times 10^{21}$ per year, as envisaged in \cite{driver},      %{driver}
values of $\sin^2 2\theta_{13}$ well below 10$^{-3}$ will become 
accessible in a few years of data taking with the UNO detector at
either of the three candidate sites.
The reach in $\sin^2 2\theta_{13}$ will thus be comparable
to that of the second phase of JHF2K based on Hyper-Kamiokande 
\cite{jhf2k}, but higher neutrino energy than in the latter experiment
will also allow to probe the sign of $\Delta m^2_{31}$. Indeed, 
switching the sign of $\Delta m^2_{31}$ effectively changes the ratio 
between the $\nu$ and $\bar{\nu}$ signals by a large factor of some 
5, 14, and 43 for $L = 1280$, 1770, and 2620 km, respectively.

     A far site corresponding to $L = 900$ km and $\theta_\nu = 11.1$
mrad is available with the original beamline of the NuMI program 
\cite{loi}, and one may assume that a water Cherenkov detector with
fiducial mass of 100 kt is built there. We estimate that in 5 years
of operation with the $\nu$ ($\bar{\nu}$) beam, this experiment will
be sensitive to $\sin^2 2\theta_{13}$ values down to 0.0024 (0.0036) 
for $\Delta m^2_{31} > 0$ ($\Delta m^2_{31} < 0$). Thus, the 
sensitivity to $\sin^2 2\theta_{13}$ will be better than in the
first phase of the JHF2K experiment \cite{jhf2k}.

     Note that unlike the 1$e$ signature, the $e\pi$ and multi-$e$ 
signatures may allow to estimate the NC background from the data: 
constraining the axes of all rings to a common point in space will 
yield the position of the primary vertex. Within errors, this should 
coincide with the reconstructed vertex of a $e^-$-induced shower, whereas 
the vertex of an unresolved $\pi^0$ shower will be displaced along the
shower direction by $\sim \lambda_\gamma$. Here, 
$\lambda_\gamma \simeq 55$ cm is the mean free path for photons in
water. This has to be compared with spatial resolution for the vertex
of a single $e$-like ring, estimated as 34 cm for 
Super-Kamiokande \cite{resolution}.                               %{resolution}
Therefore, provided that spatial resolution of UNO is sufficiently 
high, it may prove possible to directly estimate the NC background by 
analyzing the displacement of shower vertices from the reconstructed
vertex of neutrino collision.

     To conclude, we have formulated multi-ring signatures of the
oscillation \omue\ in a water Cherenkov detector, that are appropriate 
for long-baseline neutrino experiments operating at relatively high
neutrino energies $E_\nu > 2$ GeV. These emphasize the MSW matter effect
and, therefore, may allow to determine the sign of the "atmospheric"
mass-squared difference $\Delta m^2_{31}$. The NC background is
less for selected multi-ring events than for 1$e$-like events, and
may be directly estimated from the data. Sensitivity to the sign of
$\Delta m^2_{31}$ and to $\sin^2 2\theta_{13}$ has been estimated for 
a conceptual scheme in which the proposed large water Cherenkov detector,
UNO, is irradiated by a neutrino beam from Fermilab's Main Injector 
over a baseline of either 1280, 1770, or 2620 km.
Compared to the second phase of the JHF--Kamioka experiment, the 
proposed scheme may show similar sensitivity to $\sin^2 2\theta_{13}$ 
and superior sensitivity to the sign of $\Delta m^2_{31}$. We have
also discussed the physics potential of using a a water Cherenkov
detector in the NuMI program.
%  $\sin^2\Theta_{13}$
%  $\nu N \rightarrow \nu \pi^0 X$  
%  $\nu_\mu n \rightarrow \mu^- \pi^0 p$
%  $\nu N \rightarrow \nu \pi^0 \pi^0 N$
%  $\nu_e N \rightarrow e^- X$ 
%  $\nu_e N \rightarrow e^- \pi^0 X$ 
%  $\nu N \rightarrow \nu \pi^0 X$  
%  $\nu N \rightarrow \nu \pi^0 \pi^0 X$
%  $\nu_\tau N \rightarrow \tau^-X$  

\clearpage

\end{document}